\documentclass{wiley-article}

\usepackage{textgreek}
\usepackage{siunitx}
\usepackage{amsmath}
\usepackage{amssymb}
\usepackage{textgreek}
\usepackage{siunitx}
\usepackage{amsmath}
\usepackage{amssymb}
\usepackage[numbers]{natbib}
\usepackage{times}
\usepackage{wrapfig}
\usepackage{epsfig}
\usepackage{graphicx}
\usepackage{listings}

\usepackage{tikz}
\usepackage{xcolor}
\usepackage{times}

\usepackage{wrapfig}
\usepackage{epsfig}
\usepackage{graphicx}
\usepackage{listings}

\usepackage{tikz}
\usepackage{xcolor}
\usepackage{siunitx}

\usepackage{times}
\usepackage{wrapfig}
\usepackage{epsfig}
\usepackage{graphicx}
\usepackage{listings}

\usepackage{tikz}
\usepackage{xcolor}

\usepackage{times}

\usepackage{wrapfig}
\usepackage{epsfig}
\usepackage{graphicx}
\usepackage{listings}

\usepackage{tikz}
\usepackage{xcolor}
\papertype{Whitepaper}
\paperfield{PrivateKit: MIT}

\title{Apps Gone Rogue: Maintaining Personal Privacy in an Epidemic}

\author[1]{Ramesh Raskar}
\author[2]{Isabel Schunemann}
\author[3]{Rachel Barbar}
\author[1]{Kristen Vilcans}
\author[1]{Jim Gray}
\author[1]{Praneeth Vepakomma}
\author[4]{Suraj Kapa}
\author[5]{Andrea Nuzzo}
\author[6]{Rajiv Gupta} 
\author[1]{Alex Berke}
\author[1]{Dazza Greenwood}
\author[8]{Christian Keegan}
\author[2]{Shriank Kanaparti}
\author[2]{Robson Beaudry}
\author[2]{David Stansbury}
\author[2]{Beatriz Botero Arcila}
\author[9]{Rishank Kanaparti} 
\author[1]{Vitor Pamplona}
\author[1]{Francesco M Benedetti}
\author[2]{Alina Clough}
\author[7]{Riddhiman Das}
\author[2]{Kaushal Jain}
\author[1]{Khahlil Louisy}
\author[10]{Greg Nadeau}
\author[1]{Vitor Pamplona}
\author[7]{Steve Penrod}
\author[1]{Yasaman Rajaee}
\author[1]{Abhishek Singh}
\author[7]{Greg Storm}
\author[11]{John Werner}
\author[1]{Ayush Chopra}
\author[1]{Gauri Gupta}
\author[1]{Vivek Sharma}

\affil[1]{Massachusetts Institute of Technology}
\affil[2]{Harvard University}
\affil[3]{Genetic Consulting}
\affil[4]{Mayo Clinic}
\affil[5]{GSK US}
\affil[6]{Massachusetts General Hospital}
\affil[7]{TripleBlind}
\affil[8]{Westminster College}
\affil[9]{Computer Science Consulting}
\affil[10]{Public Consulting Group}
\affil[11]{Link Ventures}
\corremail{vepakom@mit.edu}

\begin{document}

\maketitle

\tableofcontents

\begin{epigraph}{-Tedros Adhanom, Director General of the WHO}
“Containment [...] must remain the top priority for all countries. With early aggressive measures countries
can stop transmission and save lives.”
\end{epigraph}
\begin{epigraph}{-Lee Su-young, Psychiatrist at Myongji Hospital, South Korea}
"[Some of my patients] were more afraid of being blamed than dying of the virus”
\end{epigraph}

\section{Introduction}
Containment, the key strategy in quickly halting an epidemic, requires rapid identification and quarantine of
the infected individuals, determination of whom they have had close contact with in the previous days and
weeks, and decontamination of locations the infected individual has visited. Achieving containment demands
accurate and timely collection of the infected individual’s location and contact history. Traditionally, this
process is labor intensive, susceptible to memory errors, and fraught with privacy concerns. With the recent
almost ubiquitous availability of smart-phones, many people carry a tool which can be utilized to quickly
identify an infected individual’s contacts during an epidemic, such as the current 2019 novel Coronavirus
(COVID-19) crisis. Unfortunately, the very same first-generation contact-tracing tools can also be – and have
been – used to expand mass surveillance, limit individual freedoms and expose the most private details about
individuals. \par
We seek to outline the different technological approaches to mobile-phone based contact-tracing to date
and elaborate on the opportunities and the risks that these technologies pose to individuals and societies.
We describe advanced security enhancing approaches that can mitigate these risks and describe trade-offs
one must make when developing and deploying any mass contact-tracing technology. Finally, we express our
belief that citizen-centric, privacy-first solutions that are open source, secure, and decentralized (such as MIT
Private Kit: Safe Paths) represent the next-generation of tools for disease containment in an epidemic or a
pandemic.
With this paper, our aim is to continue to grow the conversation regarding contact-tracing for epidemic and
pandemic containment and discuss opportunities to advance this space. We invite feedback and discussion.

\section{ The Case for Implementing Contact Tracing Technologies}

Infectious diseases spread in an exponential fashion. Containment is an effective means to slow the spread,
allowing health care systems the capacity to treat those infected. However, ‘lock down’ like containment can
also disrupt the productivity of the population, distort the markets (limiting transportation and exchange of
goods), and introduce fear and social isolation for those that are not yet infected or that have recovered from
an infection. 

\subsection{ A Timely Solution: Contact-Tracing}

Several infectious diseases have incubation periods and asymptomatic manifestation, making it difficult to
effectively measure the actual number of infected members of the population. Blanket-testing to avoid
missing asymptomatic cases, of course, is not always feasible. Another approach is contact tracing, which
involves keeping track of the possible routes of infection:

\begin{epigraph}{-World Health Organization (WHO)}
“People in close contact with someone who is infected with a virus, […], are at
higher risk of becoming infected themselves, and of potentially further infecting
others .Closely watching these contacts after exposure to an infected person will
help the contacts to get care and treatment, and will prevent further transmission
of the virus.”
\end{epigraph}
The process for contact-tracing, according to WHO, occurs in three steps:\begin{enumerate}
    \item Contact Identification: From confirmed cases, identify those the infected patient had contact
with (according to the transmission modalities of the pathogen)
\item Contact Listing: Record the possible contacts of the infected patients and contact those
individuals
\item Contact Follow-Up
\end{enumerate}
Contact tracing is a key public health response to slowing the spread and containing infectious diseases. By
mitigating the flaws of detection based solely on symptoms, contact-tracing increases the sensitivity and the
readiness of the community for an emerging epidemic. Further, contact-tracing allows citizens to relieve
burden from a community’s containment measures, as it pushes perspective infected members to isolate
themselves voluntarily as shown recently in the NYC area. \par
Finally, and most importantly, contact tracing can be quickly deployed at the first warnings of an outbreak,
but continues to be effective when disease resurgence concerns exist. Thus, following an initial epidemic
peak, contact-tracing can be an effective means to enable disease decline and avoid multiple peak periods
and disease resurgence.

\subsection{Epidemiological Impact of Application for Coronavirus Infection Contact Tracing}

Lessons from China have suggested the utility of understanding GPS localization of intersections between
known infected individuals and others in stemming infection progression. This is specifically related to the R0
(R naught) that determines how contagious an infectious disease is. R0 is a description of the average number
of people who will catch a disease from one contagious person. Ideally, a lower number will optimize
reduction of disease spread, which will facilitate time to develop a vaccine or for the disease to die out. Three
factors that define R0 are the infectious period (which is generally fixed for a given disease), the contact rate
(i.e., how many people come in contact with a contagious person), and the mode of transmission (which is
similarly fixed for a given disease). Thus, for a given disease, the most adjustable factor is the contact rate. \par
One key issue with contact rate is how to optimally allow individuals and societies to limit the contact rate.
Contact amongst uninfected individuals will not facilitate disease spread. Thus, ideally a society and/or an
individual is principally concerned with understanding the contacts an infected individual has had.
Understanding if paths have been crossed between an infected individual and any number of other individuals will allow for identifying those who have been exposed (and maybe should be tested resulting in
appropriate resource allocation or may isolate themselves in the absence of available testing). Thus, at a
societal level, this may limit the economic and public impact. \par \par  With an application that allows for users to understand potential exposure to an infected individual, and
appropriate action of the exposed individuals, it may be possible to reduce the contact rate by more rapidly
identifying cases/exposures which will remove them from the contact chain. For example, if we assume
uptake of an application amongst x\% of a population, and assuming that portion of the population responds
to known exposure by self-quarantining or pursuing texting to confirm lack of infection, the R0 will decrease
in turn by a multiple of that percentage based on the degree of mixing in the population. The reason for the
multiple decrease is R0 partially depends on the population size and density and the exact number of people
an individual may come in contact with after exposure which varies amongst individuals. Furthermore, with
an increasing number “x” in terms of user base, there will be an exponential decrease in R0 (e.g., for 100\%
use and appropriate action, R0 would be expected to fall <1 due to maximal reduction of contact rate). Thus,
for example, a 10\% uptake will have downstream impacts on individuals that person may have come in
contact by more rapid exposure/contact identification. This may eventually disrupt the contact rate with may
significantly reduce the R0 more than is accounted for by the 10\%. \par 
This ultimate effect of R0 with a 10\% use and appropriate response to data will hopefully disrupt ongoing
chains of transmission, thus effecting the mortality rate and eventually impacting the contact rate and
infection curve. However, high enough utilization could reduce contact rate to such a degree as to make the
overall R0 < 1 which would ideally lead to dying off of the infection entirely.

\section{ The Landscape of Interventions}
%\printendnotes
Almost half of the world’s population carries a device capable of GPS tracking. With this capability, location
trails—timestamped logs of an individual’s location— can be created. By comparing a user’s location trails
with those from diagnosed carriers of infectious disease, one can identify users who have been in close
proximity to the diagnosed carrier and enable contact-tracing. As the COVID-19 outbreak spreads,
governments and private actors have developed and deployed various technologies to inform citizens of
possible exposure to a pathogen. In the following, we give a brief overview over these technologies.
\\\\
\textbf{Key Terms}
We take this opportunity to define several critical terms used throughout this paper.
\begin{itemize}
    \item \underline{Users} are individuals who have not been diagnosed with an infectious disease who seek to use a
contact-tracing tool to better understand their exposure history and risk for disease.

\item \underline{Diagnosed carriers} then, refers to individuals who have had a confirmatory diagnostic test and are
known to have an infectious disease. Of note, in the setting of an epidemic in which some infected
individuals have mild or no symptoms, a subset of users will in fact be unidentified carriers. An
inherent limitation in all containment strategies is the society’s ability to identify and confirm
disease

\item \underline{Location trails} refer to the time-stamped list of GPS locations of a device, and presumably therefore,
the owner of the device.
\item Finally, we broadly speak of \underline{the government} as the entity which makes location data public and
informs those individuals who were likely in close contact with a diagnosed carrier, acknowledging that this responsibility is carried out by a different central actor in every continent, country or local
region.
\item \underline{Local businesses} refer to any private establishment such as shops, restaurants or fitness clubs as well
as community institutions like libraries and museums.
\end{itemize}

\subsection{Broadcasting}
Broadcasting refers to any method, supported by technology, by which governments publicly share locations
that diagnosed carriers have visited within the time frame of contagion. Governments broadcast these
locations through several methods. For example, Singapore updates a map with detailed information about
each COVID-19 case. South Korea sends text messages containing personal information about diagnosed
carriers to inform citizens. In the US, Nebraska and Iowa published information of where diagnosed carriers
have been through media outlets and government websites. Broadcasting methods can be an easy and fast
way for a government to quickly make public this information without the need for any data from other
citizens. It requires citizens to access the information provided and evaluate whether they may have come in
contact with a diagnosed carrier of a pathogen themselves. However, broadcasting methods risk exposing
diagnosed carriers’ identities and require exposing the locations with which the diagnosed carrier interacted,
making these places, and the businesses occupying them, susceptible to boycott, harassment, and other
punitive measures. 

\subsection{ Selective Broadcasting}
Selective broadcasting releases information about locations that diagnosed carriers have visited to a select
group, rather than the general public. For example, information might be selectively broadcast to people
within a single region of a country. Selective broadcasting requires collection of information, such as a phone
number or current location, from users in order to define the selected groups. Often, a user must sign up and
subscribe to the service, e.g., via a downloaded app. \par
Selective broadcasting operates under one of two modes: (i) The broadcaster knows the (approximate)
location of the user and sends a location specific message. Thus, user location privacy is compromised. (ii)
The broadcaster sends a message to all users, but the app displays only the messages relevant to the user’s
current location. The second approach is typically used when messages are intermittent. KatWarn, a German
government crisis app that, once downloaded and granted access to location data, notifies users within a
defined area of any major event that may impact their safety such as a natural disaster or terrorist attack.
User privacy is compromised by apps using the first mode as the broadcasting agent receives information
about the user’s location. Apps using the second mode do not have this same limitation as location data is
not reported back to the broadcaster. \par
In addition to the risk to the user’s privacy with selective broadcasting, the same risks of identification of the
diagnosed carrier and harassment of locations associated with the diagnosed carrier seen with broadcasting
apply. Further, requiring a user to sign up and subscribe risks decreased participation by possible users.

\subsection{Unicasting}
Unicasting informs only those users who have been in close contact with a diagnosed carrier. Unicasting
requires government access data, not only of diagnosed carriers, but also of every citizen who may have
crossed their path. The transmission is unique to every user. China developed a unicasting system which
shows who poses a risk of contagion. While highly effective at identifying users exposed to contagion for
containment interventions, unicasting presents a grave risk for a surveillance state and government abuse.
\subsection{Participatory Sharing }In participatory sharing, diagnosed carriers voluntarily share their location trails with the public without
prompting by a central entity, such as a government. Advantageously, with participatory sharing, diagnosed
carriers retain control of their data and presumably consent to its release. Users are required to
independently seek the information and assess their own exposure risks. However, these solutions present
challenges as it is difficult to check for fraud and abuse.

\subsection{Private Kit: Safe Paths }

\textbf{Private Kit: Safe Paths} is an MIT-led, free, open-source and privacy-first contact-tracing technology that
provides individual users information on their interaction with COVID-19, while also empowering
governments’ efforts to contain an epidemic outbreak. The solution is a ‘pull’ model where users can
download encrypted location information about carriers so the users can self-determine their likely exposure
to COVID-19 and coordinate their response with their doctor using their symptoms and personal health
history. \par

The \textbf{Private Kit: Safe Paths} solution, in its first iteration, enables individuals to log their own location. With
consent they can provide health officials with an accurate location trail once they are diagnosed positive.
Additionally, governments are equipped with a tool to redact location trails and thus broadcast location
information with privacy protection for diagnosed carriers and local businesses. In its second iteration, \textbf{Safe
Paths} provides users with information on whether they have crossed paths with a diagnosed carrier. \textbf{Safe
Paths’} ability to do so without collecting information on the user in an external cloud prevents government
surveillance. As an open-source tool, \textbf{Safe Paths} fosters public trust and utilizes experts to audit its security
and privacy features.

In the last phase of development, \textbf{Private Kit: Safe Paths} will move to a mix of participatory sharing and
unicasting, eliminating the need for a central entity while still providing a highly personalized exposure risk
assessment to users. In this third iteration, \textbf{Safe Paths} enables privacy protected participatory sharing of
location trails by diagnosed carriers and direct notification of users who have been in close proximity to a
diagnosed carrier without allowing a third party, particularly a government, to access individual location
trails.

Different technological interventions for contact-tracing pose various risks to individuals and the public. We
will discuss and compare the main challenges of deploying these technologies in the following chapter and
compare how the \textbf{Private Kit: Safe Paths} solution maximizes stakeholder value when trading off the key
constraints as compared to existing solutions.

\section{Risks and Challenges}
Risks exist for both the individual and the public with use of contact-tracing technology. The primary
challenge for these technologies, as evident from their deployment in the COVID-19 crisis, remains securing
the privacy of individuals, diagnosed carriers of a pathogen, and local businesses visited by diagnosed
carriers, while still informing users of potential contacts. Additionally, contact-tracing technologies offer
opportunities for bad actors to create fear, spread panic, perpetrate fraud, spread misinformation, or
establish a surveillance state.

\subsubsection{Privacy Risks for Diagnosed Carriers}

All containment strategies require analysis of diagnosed carrier location trails in order to identify other
individuals at risk for infection. Diagnosed carriers, therefore, are at the greatest risk of their privacy being
violated, for example, by public identification. Even when personal information is not published, these
individuals may be identified by the limited set of location data points released. When identified publicly,
diagnosed carriers often face harsh social stigma and persecution. In one example, data sent out by the South
Korean government to inform residents about the movements of those recently diagnosed with Covid-19
sparked speculations about individuals’ personal lives, from rumors of plastic surgery to infidelity and
prostitution. Online witch hunts aiming to identify diagnosed carriers create an atmosphere of fear. As
painfully articulated by the following quote, social stigma can be worse than the disease.

\begin{epigraph}{-Lee Su-young, Psychiatrist at Myongji Hospital, South Korea}
"[Some of my patients] were more afraid of being blamed than dying of the virus”
\end{epigraph}

With all currently available contact-tracing technologies, the risk for public identification of the diagnosed
carrier remains high. Further innovation is necessary to protect high risk populations.

\subsubsection{Privacy Risks for Users}Users also face privacy violations. Providing an exposure risk assessment to the user requires the user’s
location data in order to establish where the user’s path has crossed with that of a diagnosed
carrier. However, enabling access to the individual'slocation data by a third party, particularly a government,
preludes a step towards a surveillance state, as examples from the COVID-19 crisis show. In China, users
suspect an app developed to help citizens identify symptoms and their risk of carrying a pathogen spies on
them and reports personal data to the police. The Google Play store also pulled the Iranian government’s
app amid similar fears and South Korea’s app to track those in self-quarantine automatically notifies the
user's case worker if they leave their quarantine zone.

\subsubsection{Privacy Risks for Local Businesses}

Identities of cafes, shops, and other businesses visited by a diagnosed carrier may be divulged when the
carrier’s location trail is released to the public. Public association with the path of a diagnosed carrier, as
examples from China and South Korea show, damages local businesses. At a time of heightened vulnerability
due to the economic stress which often coincides with an epidemic, these businesses may suffer significant
financial hardship and possibly collapse.

\subsubsection{Privacy Risks for Non-Users} Contact-tracing technology may, at times, violate the privacy of a non-user. Users and non-users are
networked together through social relationships and environmental proximity. When a family member or
friend’s identity as a diagnosed carrier is revealed, non-users close to the diagnosed carrier may endure the
same public stigmatization and social repercussions. When a business loses customers or faces harassment
due to association with a diagnosed carrier’s location trail, its patrons and, particularly, its employees bear
the economic and social burden whether or not they are a user of contact-tracing technology. Non-users
may be further negatively affected if location trails pinpoint sensitive locations, such as military bases and
secure research laboratories.

\subsubsection{Consent and Choice}
Obtaining consent for any form of data collection and use helps manage privacy risks. Consent’s utility in
real-world settings, however, is often undermined. Language which is incomprehensible for typical users
and a lack of real choice (e.g. users must often relinquish privacy and share their data in order to receive a
service or opt not to use the service at all) severely limit the power of consent. Contact-tracing technologies
have yet to overcome the challenges associated with obtaining true consent from the user. Typically, a user
may be required to share their location with a third party in order to receive an exposure risk assessment.

\subsection{ Misinformation and Panic} During an epidemic, complex and quickly evolving data must be accurately conveyed to and understood by
the entire public, including individuals with low health literacy. Serious harm, including heightened alarm
among the public, may result from failure to appropriately communicate health risks. Contact-tracing
technologies have potential to introduce misinformation and cause panic. For example, if users receive an
alert about a possible contact location without appropriate information and understanding of the exposure
time frame, some users will inaccurately conclude they are at high risk. Even when information regarding
both location and time is provided to users, if the magnitude of the risk cannot be easily comprehended, an
atmosphere of fear or a run on the medical system may be provoked.

\subsection{Risky Behavior}
Feeling a false sense of safety at having not received a notification of exposure, some users may
underestimate their risk for disease. Users who no longer perceive a significant risk may be less likely to
engage in other forms of disease prevention, such as social distancing. A false sense of safety may occur
when the limitations of contact-tracing technology within a community are not clearly communicated to the
public. 
\subsection{ Fraud and Abuse}
Technological interventions in human crises are often targeted for fraud and abuse. In South Korea,
fraudsters quickly began blackmailing local merchants and demanding ransoms to not (falsely) report
themselves as sick and having visited the business. Additionally, bad actors may force individuals to provide
their location data for purposes other than disease containment, such as for immigration or police purposes.
Fear of such abuse may prevent a contact-tracing system meant to help save lives from being adopted.
\subsection{ Security of Information}
Hacking lingers as a serious risk for all data-gathering technologies with sensitive information, like health
status and location. Hackers have successfully infiltrated apps and services collecting sensitive information
before, with 92 million accounts from the genealogy and DNA testing service MyHeritage hacked in
2017. Data security must lie at the center of every effort to use location data for contact-tracing and
containment.
\subsection{Equity and Socioeconomic Factors}
Ensuring equity and social justice challenges many technologies, including contact-tracing. If participation
requires ownership of a smartphone, some people, often those most vulnerable, the elderly, the homeless,
and those living in lower-income countries, will not be able to access the technology. A lack of access to
devices among vulnerable populations will remain a significant challenge for contact-tracing technology in
the near future. Avoidance by the public may impact any business identified on a diagnosed carrier’s location
trail, but reduced hours or job loss hurt lower-income service workers most. Finally, abuse of data collection and violations of user privacy are inflicted more often upon those who are already most vulnerable to
government surveillance.

\section{ Mapping Technological Interventions with Risks}
In the following table, the various contact-tracing technological approaches are mapped against the reviewed
risks and challenges. 
\subsection{ The Utility-Privacy Trade-Off}
The inverse relationship between accuracy of the provided risk assessment and user privacy for contact tracing technologies necessitates compromise by the user community. The core trade-off between utility and
user privacy, diagrammed below, illustrates this and highlights the potential of Private Kit: Safe Paths to
fundamentally alter this relationship. 

\begin{figure}[htp]
  \begin{center}
    \includegraphics[width=15cm]{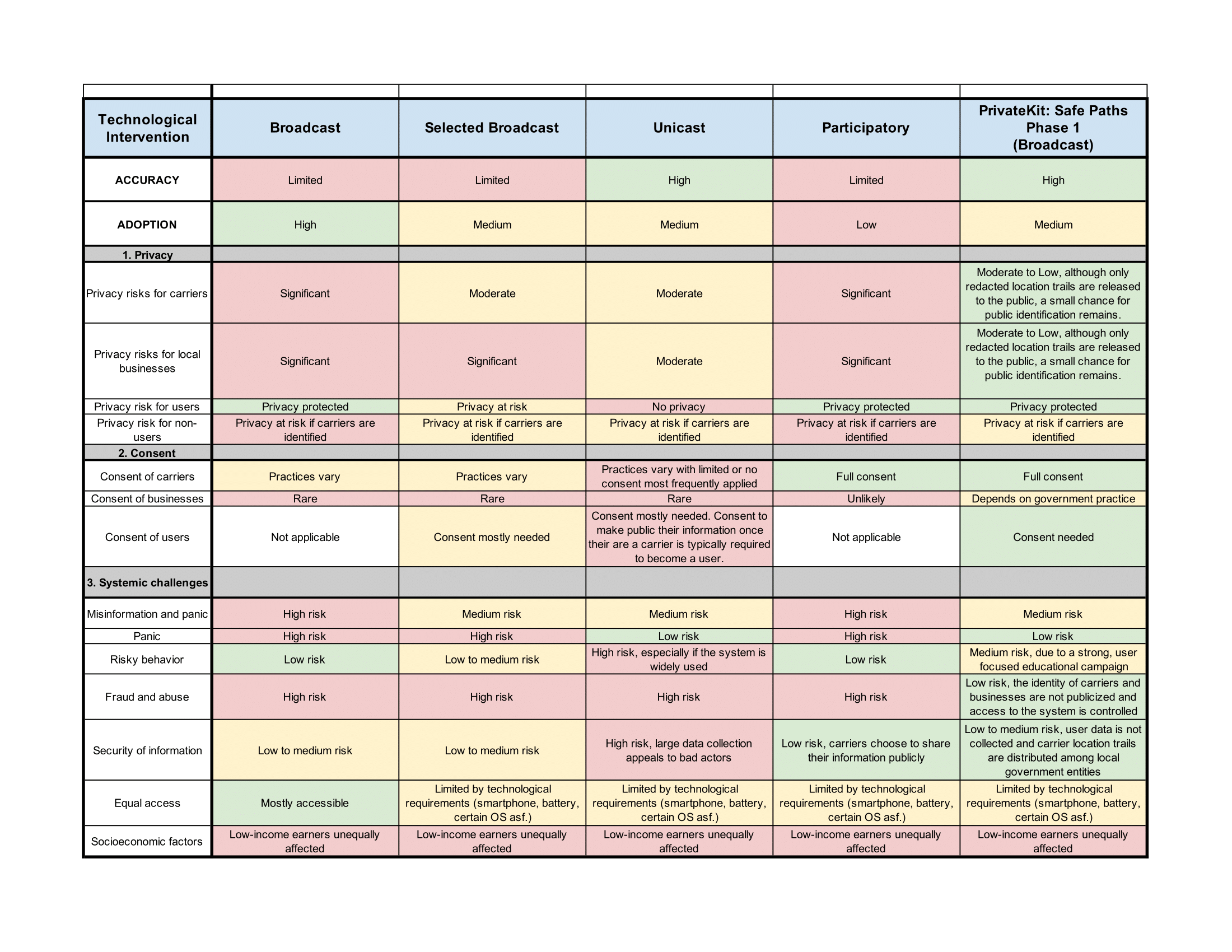}
    \caption{The various contact-tracing technological approaches are mapped against the reviewed risks and challenges.}
    \end{center}
    \label{fig:galaxy}
\end{figure}

\begin{figure}[htp]
    \begin{center}
        \includegraphics[width=6cm]{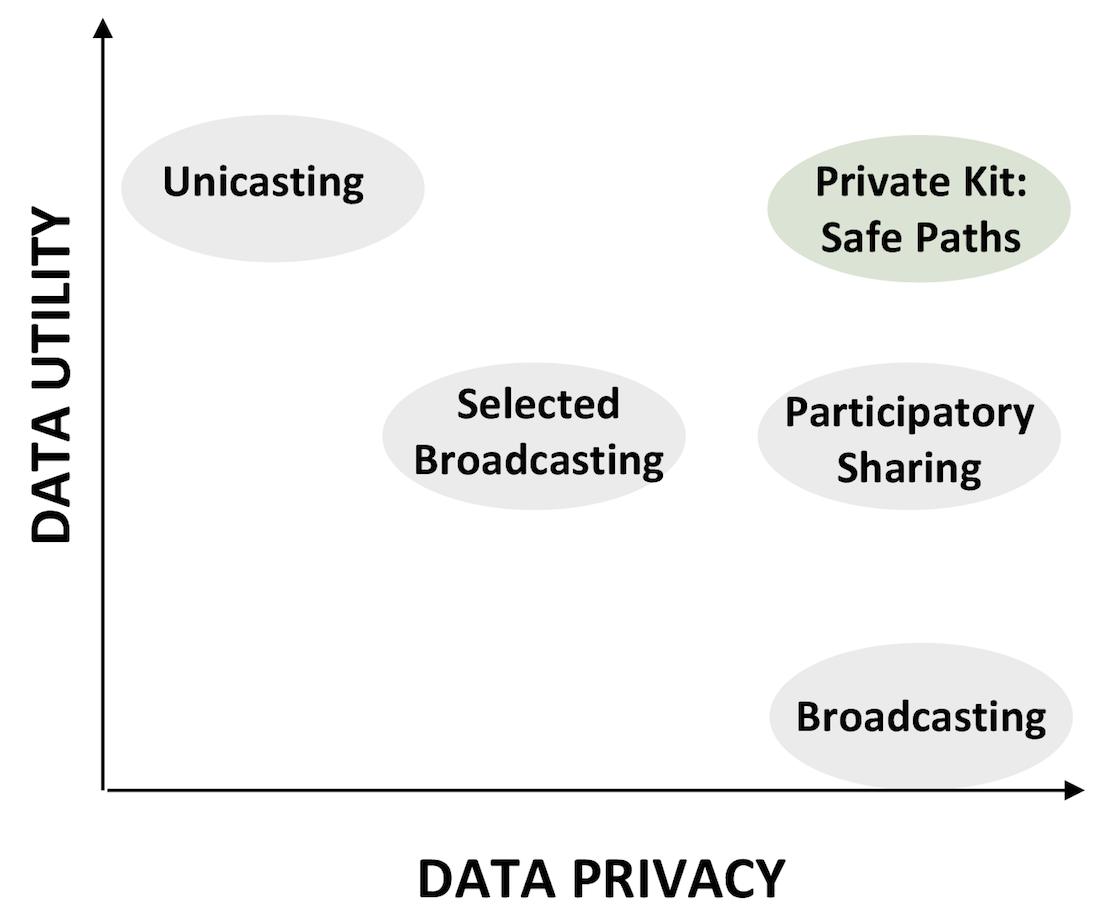}
        \caption{Unicasting, Broadcasting, Selected Broadcasting and Private Kit: Safe Paths plotted in terms of the trade-off of privacy and utility.}
    \end{center}

    \label{fig:galaxy}
\end{figure}
\section{Discussion of Risks, Mitigation and Trade-offs}

Deploying any form contact-tracing technology requires contemplation of several risks outlined in the prior
analysis. Mitigation of these risks depends on thoughtful consideration of the trade-offs inherent to contacttracing technology and containment strategies. In the following, we review decisions required for these
trade-offs and best approaches for risk mitigation. 

\subsection{ Privacy of Diagnosed Carriers}
Data must be collected from diagnosed carriers to facilitate containment of an epidemic. However, both data
collection and release of that information to identified contacts may violate the diagnosed carrier’s
privacy. As the most vulnerable stakeholder in the containment strategy, several efforts must be undertaken
to protect the diagnosed carrier’s privacy to the highest degree possible. Limiting the publicly published data
helps protect the known carrier’s identity from the public. To date, with the exception of participatory sharing
models, the diagnosed carrier’s data must be shared with a third-party entity, requiring the carrier to
relinquish at least some control over their data. Ending the need for third party involvement would represent
an immense step forward in privacy protection for diagnosed carriers. Access and usage of the data by an
entity, mostly governments, should be limited and highly regulated. Harsh penalties for the abuse of such
data should be established. Obtaining true user consent further protects diagnosed carriers. Not all
approaches in use today require consent to share personal data. Particularly in non-democratic regimes,
diagnosed carriers may be unable to deny consent. In other instances, all users must consent to share their
data in order to be informed of their own exposure risk. We believe no one should be obligated to share their
personal information. Time limited storage of location trails further protects the privacy of diagnosed
carriers. Finally, using an open-source approach to create an app fosters trust in the app’s privacy protection
capabilities, as independent experts and media can access and evaluate the source code.
\subsection{ Privacy of Local Businesses}

Containment of an epidemic requires publication of sites of known exposure to a diagnosed carrier to the
public. Yet doing so risks harassment of local businesses at these sites. Providing broader location data may
better protect the privacy of a local business, but also affects the accuracy of the risk assessment. Broad
location data, such as notice of a 100x100m area into which a diagnosed carrier sojourned, may still identify
a business. Any contact-tracing approach must balance the public health benefit of disease containment
against the threat of economic hardship for local businesses connected to the epidemic. \par
There is no easy answer to this trade-off as any choice impacts utility of the technology and risks affecting
the viability of the business. Evaluating the risk versus benefit of location data release should occur on a
case-by-case basis. The time frame of possible contagion must be released so the users may understand the
limits of the exposure risk. Critically, the entity publishing the location data should consult with the local
business and inform the business of any decision before the public is notified.

\subsection{Access and Inclusion}
Issues of access and inclusion are not easily resolved by contract-tracing technology. Limited access to a
device capable of utilizing contact-tracing technology and difficulty understanding and acting on the provided
risk assessment overly affect the more marginalized of our societies. However, containing an epidemic
outbreak quickly benefits everyone within a community. Implementation of contact-tracing technology
within a community, even with unequal access, may increase the safety of all. The development of a simple
GPS device that can share location trails may be a medium-term solution to some accessibility concerns,
particularly in countries with limited smartphone penetration. Additionally, some form of access to information about a possible contagion must be made available to those without a smartphone and all
information should be presented in a way that accounts for variation in health literacy among users.

\subsection{ Misinformation and Risky Behavior}The spread of misinformation cultivates instability and uncertainty during a crisis. Release of information on
the spread of a pathogen to the public invites public speculation and fear-mongering and manipulation by
bad actors. A false sense of safety for users may increase alongside increased efficiency of contact-tracing
technology. Entities providing contact-tracing technology are also at risk to introduce error within the release
information, despite best intentions. At this time, no strategies exist to eliminate these risks; however, such
risks can be mitigated through educational outreach efforts and engagement with key stakeholders. 
\subsection{Security of Information}
Storage of sensitive information invites attack by hackers. Trade-offs must be made in order to mitigate this
risk. Only anonymized, redacted, and aggregated sensitive information should be stored. Use of a distributed
network, rather than a central server, makes hacking less attractive, but requires providing security to
multiple sites. In the long term, the safest way to store location data will be in an encrypted database
inaccessible to all, including the government. Time limitations on data storage also work well to secure
information and should be implemented in contact-tracing technology. During an epidemic outbreak, the
appropriate amount of time for data storage equals the time during which a diagnosed carrier could have
possibly infected another individual. For COVID-19, this time frame is set to be 14 to 37 days. Deleting data
after such a short period, particularly during an outbreak of a poorly understood pathogen has risks.
However, we feel this trade-off should be made for data security and user privacy.
\section{Conclusion}Our ability to accurately trace contacts of individuals diagnosed with a pathogen and notify others who may
have been exposed has never been greater. Real risks exist, though, thus care must be addressed in the
design of the solution to prevent abuse and mass surveillance. As a beginning to the discussion of how to
develop and deploy contact-tracing technologies in a manner which best protects the privacy and data
security of its users, we have reviewed various technological methods for contact-tracing and have discussed
the risks to both individuals and societies. \textbf{PrivateKit: Safe Paths} eliminates the risk of government
surveillance. It draws on the advantages from several models of contact-tracing technology while better
mitigating the challenges posed by use of such technology. We have presented a discussion of precautions
which should be taken and trade-offs which will need to be made. We invite feedback and discussion on this
whitepaper.

\section{Acknowledgements}
We would like to acknowledge Amandeep Gill of the International Digital Health \& AI Research Collaborative
(I-DAIR), Bernardo Mariano Jr of the World Health Organization (WHO), and Don Rucker of the U.S.
Department of Health and Human Services (HHS) for their mentorship in advancing contact-tracing solutions.

\end{document}